\documentclass[11pt, letterpaper]{mystyle}
\usepackage{graphicx} 
\usepackage{caption}
\usepackage{subcaption}
\usepackage{todonotes}
\usepackage{natbib}
\usepackage{amsmath}
\usepackage{url}
\usepackage{float}
\usepackage{multirow}
\usepackage{booktabs}
\usepackage{algorithm}
\usepackage{algorithmic}
\usepackage{tcolorbox}

\title{Generating Public Health Responses using Survey-Augmented Large Language Models}

\runningtitle{Generating Public Health Responses using Survey-Augmented Large Language Models}

\author[1]{Leonardo Marciaga$^*$}
\author[2]{Thuyen Pham$^*$}
\author[3]{Julia Rezvani$^*$}
\author[4]{Alina Hyk$^*$}
\author[5]{Chunyang Liao$^\dagger$}
\author[6]{Konstantinos Mitsopoulos$^\dagger$}
\author[7]{Raffaele Vardavas$^\ddagger$}

\affil[1]{Illinois Institute of Technology}
\affil[2]{University of Massachusetts Amherst}
\affil[3]{Portland State University}
\affil[4]{Oregon State University}
\affil[5]{University of California Los Angeles}
\affil[6]{Johns Hopkins University}
\affil[7]{Causal Paths Analytics LLC}

\correspondingauthor{Leonardo Marciaga: (\texttt{lmarciaga@hawk.illinoistech.edu})}

\begin{document}

\begin{tcolorbox}[titlebox] 
\begin{abstract}
{\centering\section*{Abstract}}
Epidemiological models often rely on survey data to represent how individuals make health-related decisions, such as whether to vaccinate or adopt protective behaviors. However, repeated large-scale surveys are costly, time-consuming, and limited in the range of scenarios they can capture. In this work, we investigate whether large language models (LLMs) can generate synthetic survey responses that reproduce patterns observed in real populations. Using longitudinal data from the FluPaths surveys, we first identify groups associated with broadly positive or negative attitudes toward vaccination through clustering analysis. 
We then evaluate several LLMs using a cluster-informed prompting approach to generate synthetic survey responses across multiple epidemic waves. Across models, the synthetic data generally reproduce the distributions of demographic characteristics, vaccination-related beliefs, risk perceptions, and health behaviors observed in the survey data. However, they are less successful at capturing how these factors vary together within respondents. Some models reproduce group-level vaccination trends more reliably than others, although performance varies across waves. We also trained a classifier to distinguish real from synthetic records and found that the generated responses remained identifiable as synthetic. Overall, our findings suggest that LLM-generated survey data may provide a useful tool for exploratory data augmentation and we hope that it could support agent-based epidemic modeling approaches. However, the generated data should not be treated as a substitute for human survey data without further methodological improvements and validation.
\end{abstract}
\end{tcolorbox}

\maketitle

\vspace{15px}
\section{Introduction}

Human behavior is a critical determinant of epidemic outcomes. During infectious disease outbreaks, individual decisions, such as whether to vaccinate, adopt protective measures, or comply with public health guidelines, can fundamentally alter disease trajectories and the effectiveness of interventions. The COVID-19 pandemic demonstrated how population-level behavioral responses to evolving risks and policies shape transmission dynamics, often in ways that challenge traditional epidemiological models that treat behavior as static or exogenous.

Surveys provide empirical windows into population beliefs, attitudes, risk perceptions, and behavioral intentions during public health emergencies. They have been instrumental in tracking vaccine hesitancy, adoption of non-pharmaceutical interventions (NPIs), trust in institutions, and shifting risk perceptions as epidemics unfold. However, traditional survey-based approaches face significant constraints that limit their utility for real-time epidemic modeling and policy planning. First, conducting large-scale, repeated surveys requires substantial resources for participant recruitment, data collection, and processing, creating prohibitively high costs that constrain sample sizes, survey frequency, and questionnaire length. Second, human cognitive capacity imposes limits on the complexity of questions that can be reliably answered. Respondents struggle with compound conditional elements, multi-way interactions between variables, or scenarios requiring simultaneous consideration of multiple information sources. Third, the time lag between data collection and analysis can delay critical insights during fast-evolving public health crises when timely behavioral intelligence is most needed.

Recent advances in large language models (LLMs) offer a transformative approach to addressing these limitations. LLMs, trained on massive text corpora, exhibit sophisticated capabilities in reasoning, pattern recognition, and generating human-like responses to complex scenarios. When appropriately trained or prompted with structured survey data, LLMs can potentially serve as ``digital twins'' or synthetic respondents that replicate diverse demographic profiles and behavioral patterns, enabling scalable simulations that augment traditional data collection methods.

The promise of LLM-based digital twins for survey research has generated substantial recent interest across marketing science, computational social science, and artificial intelligence. \cite{toubia2025database} introduced the Twin-2K-500 dataset, surveying over 2,000 individuals across 500+ questions covering demographics, psychological traits, economic preferences, and behavioral experiments. Their work established a critical benchmark for validating digital twin methodologies, demonstrating that LLM-based personas could achieve 72\% average accuracy on holdout questions (88\% relative to test-retest accuracy). Similarly, \cite{park2024generative} presented a novel agent architecture that simulates 1,052 real individuals by applying LLMs to two-hour qualitative interviews about their lives, achieving 85\% accuracy relative to test-retest reliability on the General Social Survey and comparable performance in predicting personality traits and experimental outcomes. Multiple studies have explored LLMs as virtual survey respondents across diverse domains, from sociodemographic simulation \citep{chen2025large} to economic experiments \citep{horton2023large, aher2023using} and market research \citep{goli2024frontiers}. These studies demonstrate the technical feasibility of creating individual-level digital replicas that can answer novel questions with meaningful accuracy.

However, important uncertainties remain about the reliability and appropriate use cases for LLM-based survey augmentation. \cite{bisbee2024synthetic} provided a cautionary perspective in their analysis of using LLMs as ``synthetic replacements'' for human survey data, demonstrating that ChatGPT-generated responses on political feeling thermometers showed systematic biases and substantially reduced variance compared to matched human respondents from the American National Election Studies. Research by \cite{brucks2025digital} revealed that LLM responses can be influenced by prompt architecture, such as the ordering or labeling of response options. \cite{santurkar2023whose} found that LLMs exhibit strong systematic ordering biases and consistently high entropy in survey responses, independent of question content, making it challenging to draw robust conclusions about their inherent properties. Multiple studies \citep{motoki2024more, li2024frontiers} suggested that LLMs tend to express opinions that systematically differ from human population distributions, often reflecting training data biases toward overrepresented demographic groups. Recent work by \cite{peng2025digital} provided a sobering assessment: across 164 outcomes in 19 pre-registered studies, digital twins showed only modest correlation with human responses (approximately 0.2 on average) and exhibited less variability than human responses, suggesting they may smooth over important individual differences.

Addressing these limitations requires careful methodological approaches. \cite{huang2025how} proposed an uncertainty quantification framework that converts LLM-simulated responses into reliable confidence sets for population parameters, adaptively selecting simulation sample sizes to achieve nominal coverage despite distribution shift between synthetic and real populations. Their work provides a quantitative measure of LLM ``simulation fidelity'' by determining the effective human population size that an LLM can represent. \cite{jansen2023employing} argued that while LLMs show promise for addressing some survey research challenges (question wording, response bias), they must be used in conjunction with other methods and cannot yet correct for sampling or nonresponse bias.

\subsection{Motivation and Research Context}

\subsubsection{The COVIDPaths Project and Its Limitations}

This work emerges from our ongoing NIAID R01-funded COVIDPaths project, which constructs an agent-based model (ABM) addressing the joint behavioral and disease dynamics of COVID-19 and seasonal influenza. Our modeling framework is informed by an 8-wave longitudinal survey fielded to the probability-sampled American Life Panel (ALP), comprising nationally representative respondents across multiple epidemic waves from 2016 through 2024. The FluPaths and COVIDPaths datasets provide rich, temporally-resolved measurements of vaccination behaviors, risk perceptions, protective measures, and social network influences across the evolution of two major respiratory disease epidemics.

Our current approach involves an iterative feedback loop: survey data inform behavioral parameters in the simulation model, while the structural requirements of the simulation model shape survey design and question selection. This tight integration has proven scientifically productive, enabling us to capture evolving behavioral responses to changing epidemic conditions and policy interventions. However, this approach faces three fundamental limitations that motivated the current investigation:

\textbf{Cost and Scalability Constraints.} Each wave of our longitudinal survey requires substantial financial resources for panel maintenance, participant compensation, survey programming, and data processing. These costs directly limit sample sizes (constraining our ability to capture rare subpopulations), survey frequency (reducing temporal resolution during rapidly changing epidemic phases), and questionnaire length (forcing difficult tradeoffs between breadth and depth of measurement). For a typical wave in our study, costs can reach tens of thousands of dollars, making it infeasible to conduct the frequent, large-scale surveys that would ideally inform a continuously updated epidemic model.

\textbf{Cognitive Complexity Limitations.} Agent-based epidemic models benefit from detailed behavioral response functions that capture how individuals react to complex, interacting factors: disease prevalence, vaccination rates, policy stringency, social influence, personal health status, and subjective risk perceptions. Ideally, we would elicit how protective behaviors vary across many specific combinations of these factors. However, human respondents face cognitive constraints in answering such complex questions. Asking someone to estimate ``How likely would you be to get vaccinated if 30\% of people in your community were vaccinated, COVID-19 hospitalizations were at 50 per 100,000, and mask mandates were in effect?'' requires integrating multiple information sources and hypothetical scenarios, which is a cognitively demanding task prone to satisficing, response fatigue, and unreliable answers. We must therefore limit ourselves to simpler, lower-dimensional questions that sacrifice the precision our models could utilize.

\textbf{Counterfactual and Scenario Exploration Limitations.} Epidemic preparedness and policy optimization require exploring scenarios that may not have occurred in observed data. What would vaccination uptake look like under different vaccine efficacy profiles? How would protective behaviors change with different public health messaging strategies? How might behavior differ in hypothetical future epidemic waves with novel characteristics? While we can ask human respondents some hypothetical questions, we are severely limited in the number and complexity of counterfactual scenarios we can reliably explore through traditional surveys.

\subsubsection{Benefits of LLM Augmentation}

These limitations create a compelling opportunity for LLM-based survey augmentation specifically adapted to epidemic modeling contexts. Unlike general consumer research or political polling applications, epidemiological surveys have unique characteristics that both challenge and potentially facilitate LLM augmentation approaches:

\textbf{Temporal Structure and Learning Opportunities.} Our longitudinal data spans multiple epidemic waves, capturing how the same individuals' beliefs and behaviors evolve in response to changing disease dynamics, policies, and information environments. This temporal richness provides opportunities to train or fine-tune LLMs not just on static cross-sectional patterns, but on dynamic behavioral trajectories. Understanding whether someone's vaccination behavior changed from one wave to another, and what factors predicted that change, offers signal beyond single-timepoint responses.

\textbf{Domain-Specific Context and Constraints.} Unlike open-ended consumer preferences, public health behaviors operate within structured decision contexts: vaccination involves specific attributes (efficacy, safety, timing), protective measures involve concrete actions (masking, distancing, ventilation), and risk perceptions respond to measurable epidemiological indicators (case rates, hospitalizations, deaths). This domain structure may help constrain LLM generations toward realistic responses compared to more abstract opinion questions.

\textbf{Complementary Strengths with Agent-Based Models.} Our ultimate goal is not to replace human surveys entirely, but to create a multi-loop system where (1) expensive, high-quality human survey data calibrates and validates LLM-based synthetic respondents, (2) synthetic respondents answer more numerous, complex, and counterfactual questions to inform ABM parameters, and (3) ABM simulations identify which behavioral relationships matter most, focusing subsequent human data collection on high-value measurements. This hybrid approach could amplify the value of each human response collected. Recent work on LLM-augmented agent-based modeling \citep{gao2024large, williams2023epidemic} have demonstrated promise for social simulations, including epidemic spread where LLM agents have been shown capable of replicating complex phenomena such as multi-peak outbreak patterns. However, \cite{demooij2023framework} noted that the current literature lacks unified solutions that both scale to population-level epidemic simulations and support deliberative agents with realistic human behavioral complexity, which is precisely the gap our hybrid approach aims to address. Related efforts \citep{mitsopoulos2023psychologically} have proposed combining cognitive architectures and large language models to support scalable agents with cognitively-constrained decision-making capabilities.

\subsubsection{Research Objectives and Approach}

In this work, we provide findings from an initial investigation on whether and how LLMs can generate synthetic preventive health behaviors from longitudinal epidemic survey data from our ALP dataset. Specifically, we aim to:

\begin{enumerate}
    \item \textbf{Assess LLM capability to capture behavioral patterns} in epidemic survey data, including cross-sectional distributions across multiple waves.
    
    \item \textbf{Develop and evaluate augmentation methodologies} that leverage clustering analysis and cluster-informed prompting to improve synthetic data quality and alignment with human responses.
    
    \item \textbf{Compare LLM performance across models and prompting strategies}, examining which approaches best balance population-level distributional fidelity and computational efficiency for our epidemic modeling application.
    
    \item \textbf{Establish practical validation frameworks} for epidemic-specific survey augmentation that assess whether synthetic data preserve key behavioral relationships needed for simulation modeling.
\end{enumerate}

Our approach differs from recent digital twin work in several important ways. First, we focus specifically on public health behavior during epidemics rather than general consumer or social attitudes, recognizing that domain-specific patterns and constraints may affect LLM performance. Second, we leverage longitudinal data spanning actual epidemic waves rather than cross-sectional snapshots, allowing us to examine whether LLMs can capture temporal behavioral dynamics. Third, our goal is explicitly to augment, not replace, human survey data for the specific purpose of informing agent-based epidemic models, rather than to create comprehensive digital replicas of individuals. Recent comprehensive surveys on LLM-empowered agent-based modeling \citep{gao2024large, gonzalezcastro2024llms} emphasize the transformative potential of integrating LLMs into simulation systems while highlighting persistent challenges in environment perception, human alignment, action generation, and evaluation. These challenges that are particularly salient in the epidemic modeling context where behavioral fidelity directly affects population health outcomes.

We approach this investigation with appropriate scientific skepticism, informed by recent findings about limitations of current LLM-based digital twin approaches. We recognize that modest correlations between synthetic and human responses at the individual level may still provide value for population-level modeling if distributional properties and key behavioral relationships are preserved. Conversely, even relatively high accuracy on holdout questions may be insufficient if LLMs fail to capture the heterogeneity and variability that drive epidemic dynamics.

The remainder of this paper proceeds as follows: Section 2 reviews related work on LLM-based survey simulation and digital twins. Section 3 describes the FluPaths and COVIDPaths datasets. 
Section 4 presents our cluster-informed prompting methodology and evaluation approach.
Section 5 reports results on synthetic data quality across multiple LLMs and evaluation metrics. Section 6 discusses findings, limitations, and implications for integrating LLM-augmented surveys into epidemic modeling workflows.

\subsubsection{Broader Implications}

Beyond our immediate application to epidemic modeling, this work contributes to the broader scientific conversation about appropriate use cases and limitations of LLM-based survey augmentation. The public health domain provides a valuable testbed because of several reasons: (1) decisions informed by behavioral models affect population health outcomes, demanding rigorous validation of synthetic data quality; (2) unlike predicting future behaviors, we have actual observed behaviors across multiple epidemic waves to serve as validation benchmarks; (3) health behaviors involve clearer decision attributes and outcomes compared to many consumer or political preference domains; and (4) repeated measurements allow testing whether LLMs can capture not just static patterns but behavioral change over time.

Our findings will inform not only epidemic modeling but the broader question of when, how, and with what precautions researchers can responsibly leverage LLM capabilities to enhance rather than replace traditional data collection methods.

\section{Related Work}
Epidemiological models have traditionally incorporated human behavior through fixed parameters or simple assumptions, but recent studies have explored machine learning approaches to predict behaviors from survey data, such as using automated machine learning on cohort studies to forecast health outcomes \citep{thirunavukarasu2023}. The high costs of conducting large-scale surveys in public health, including recruitment and processing, have prompted alternatives like LLMs to analyze existing data more efficiently. LLMs have been applied to extract insights from open-text responses in global health surveys, achieving high accuracy in categorizing reasons for vaccine non-access with few-shot prompting or fine-tuning \citep{burstein2024}. Similarly, LLMs have been used to classify public stances on vaccination from social media discourse, outperforming traditional sentiment analysis tools in detecting minority views \citep{espinosa2024}.

LLMs have shown potential in generating synthetic survey responses to simulate human opinions and beliefs. For instance, LLMs can predict associations among human attitudes, recreating correlations even across dissimilar topics by capturing latent structures in belief systems \citep{ma2025}. In survey research, LLMs have been employed to generate responses to items, addressing issues like question wording and response bias, though they require integration with other methods to handle sampling limitations \citep{jansen2023employing}. This extends to creating virtual populations for predicting survey outcomes without additional training data, with models like GPT-4o and Claude achieving competitive accuracy compared to traditional algorithms \citep{sinacola2025}.

In public health contexts, LLMs augmented with survey data have facilitated scalable understanding of discourse, such as gauging opinions on health policies through social media analysis \citep{espinosa2024}. Synthetic data generation using LLMs has also been explored in healthcare to address data scarcity, including in rare disease research where models create privacy-preserving datasets \citep{goyal2023synthetic}. Furthermore, LLMs have replicated human-like responses in surveys, enabling simulations of attitudes in response to events, though performance varies by model and context \citep{qu2024performance}.

Despite these advances, LLMs exhibit biases that affect their reliability in replicating human opinions. Training data often leads to over-representation of certain groups, resulting in reduced accuracy for underrepresented demographics, such as non-Western populations or specific religious groups \citep{sinacola2025}. LLMs tend to capture overall trends but struggle with individual variations, showing cultural, linguistic, and demographic biases that limit global applicability in public opinion simulation \citep{qu2024performance}. Claims that synthetic data can amplify underrepresented groups are challenged, as biases in training data make reliable representation statistically unlikely without targeted interventions \citep{gallegos2024bias}.
In our work we give emphasis in the pre-processing step and generating data over multiple epidemic waves. 
We also presented preliminary results at the AAAI-26 Student Abstract track \citep{aaai26}, where the focus was on comparing LLM performance across models using distributional similarity and vaccination-rate reproduction as evaluation criteria. The present paper extends that work with a fuller methodology description, cross-wave analysis, and cluster-targeted prompting.

\section{Data Overview}

In this section, we introduce the FluPaths and COVIDPaths datasets, collected by the RAND Corporation through the American Life Panel (ALP), a nationally representative internet panel of U.S. adults aged 18 and older. FluPaths and COVIDPaths are longitudinal studies that examine respondents' beliefs and behaviors across time regarding influenza and COVID-19 on a wide variety of topics, such as healthcare decisions, risk perceptions, and social network influences. These surveys comprised of waves, one for each time the surveys were accepting responses from participants. Our current analysis uses the FluPaths dataset as a proof-of-concept, with COVIDPaths extension left for future work.



FluPaths is a longitudinal survey conducted from Fall 2016 to Spring 2020, examining influenza-related beliefs and behaviors. It comprises eight waves, where odd-numbered waves were fielded in the Fall seasons and even-numbered waves in the Spring seasons. The baseline, Wave 1, was conducted in Fall 2016 and provided initial measurements of variables of interest, such as health behaviors and risk perceptions. Waves 2 to 7 were collected between 2017 and 2019, allowing for the continued measurements of these variables as well as expanding the variety of topics asked in the survey. Wave 8, fielded during the early phase of the COVID-19 pandemic, was partitioned into two waves, 8a and 8b. The former, conducted in Spring 2020, incorporated early responses and perceptions regarding the COVID-19 pandemic, while the latter, administered in Summer 2020, captured evolving behaviors and beliefs as the pandemic escalated.


COVIDPaths is an extension of FluPaths after the start of the COVID-19 pandemic. COVIDPaths surveys are also fielded twice annually, with odd-numbered waves conducted in Winter and even-numbered waves in Summer. It comprises Waves 9 to 16, where Wave 9 was conducted in Winter 2022, and Wave 16 in Summer 2025. COVIDPaths extends the focus on influenza from FluPaths to include behaviors and beliefs surrounding COVID-19, such as vaccination, adoption of non-pharmaceutical interventions (NPIs) such as masking and social distancing, and risk and policy perceptions. 



A particular challenge is the evolution of the survey instruments across. Because this survey had to adapt to emerging issues in the rapidly-evolving public health context, both the content and format of questions often change across waves. Therefore, analysis of each wave had to be done on a case-by-case basis, as we detail in Section~\ref{Sec: method}.




\section{Methodology}
\label{Sec: method}

Building on the cluster-informed prompting approach proposed by \cite{unipredict2024}, our methodology integrates additional information derived from clustering analysis of individual survey waves. 
We demonstrate this process across four waves of the FluPaths dataset, specifically Waves 1, 3, 5, and 7, and describe in detail both our clustering procedure and the prompting strategy used. 
Our methodology consists of three steps, which are depicted in Figures \ref{fig:step1}-\ref{fig:step3} and described in the following sections.

\subsection*{Step 1: Data preparation and generation of cluster descriptions}
\begin{figure}[!ht]
\centering
\includegraphics[width=\linewidth]{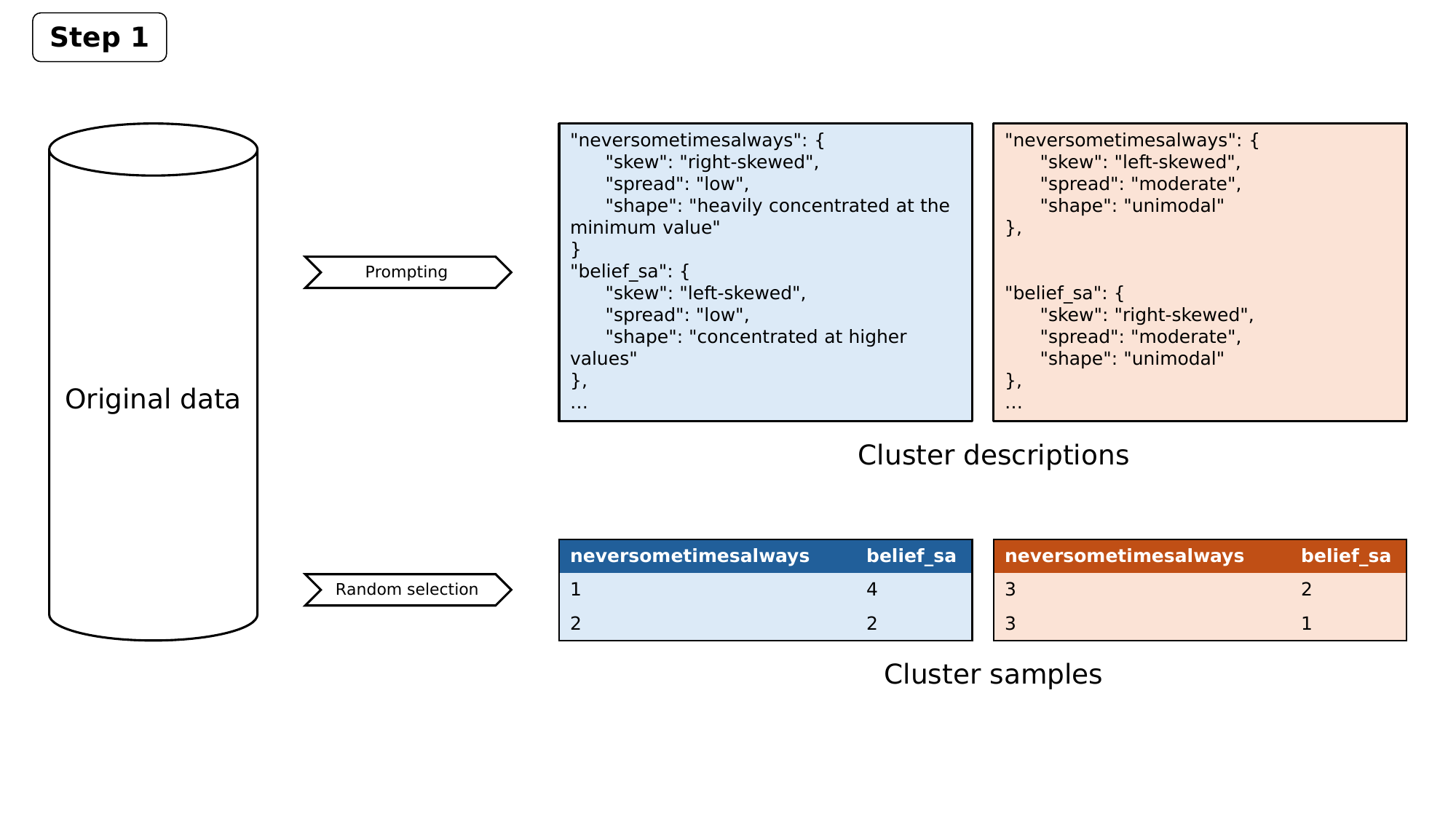}
\caption{Step 1: Data preparation and generation of cluster descriptions. An example of cluster descriptions and cluster samples.}
\label{fig:step1}
\end{figure}

We start by cleaning missing data and selecting ego-related questions from the dataset. Then, we apply clustering analysis on the selected variables (questions). 
For each wave, we select three variables corresponding to question items that are the most explicative of individual trajectories across waves and the demographic attributes sex and age. 
We implement K-means clustering on Wave 1 and Wave 7, and spectral clustering on Wave 3 and Wave 5, as those methods give better results than other clustering techniques. 
We recognize two major clusters in each wave, corresponding to generally positive and generally negative attitudes towards vaccination, respectively.
We then select 4 random samples from each of the two clusters in each wave. 
For Waves 3, 5, and 7, we also include the vaccination outcome variable not present in Wave 1. 
Finally, we prompt the LLM with summary statistics of the selected variables to obtain natural-language descriptions of the distributions of these variables in each cluster. We ask the LLM to provide the ``skewness'', ``spread'' and ``shape'' of the distribution of each variable in purely qualitative terms.
We depict Step 1 and show an example of cluster descriptions and cluster samples in Figure \ref{fig:step1}.

\subsection*{Step 2: Formatted metadata generation}
\begin{figure}[!htbp]
\centering
\includegraphics[width=\linewidth]{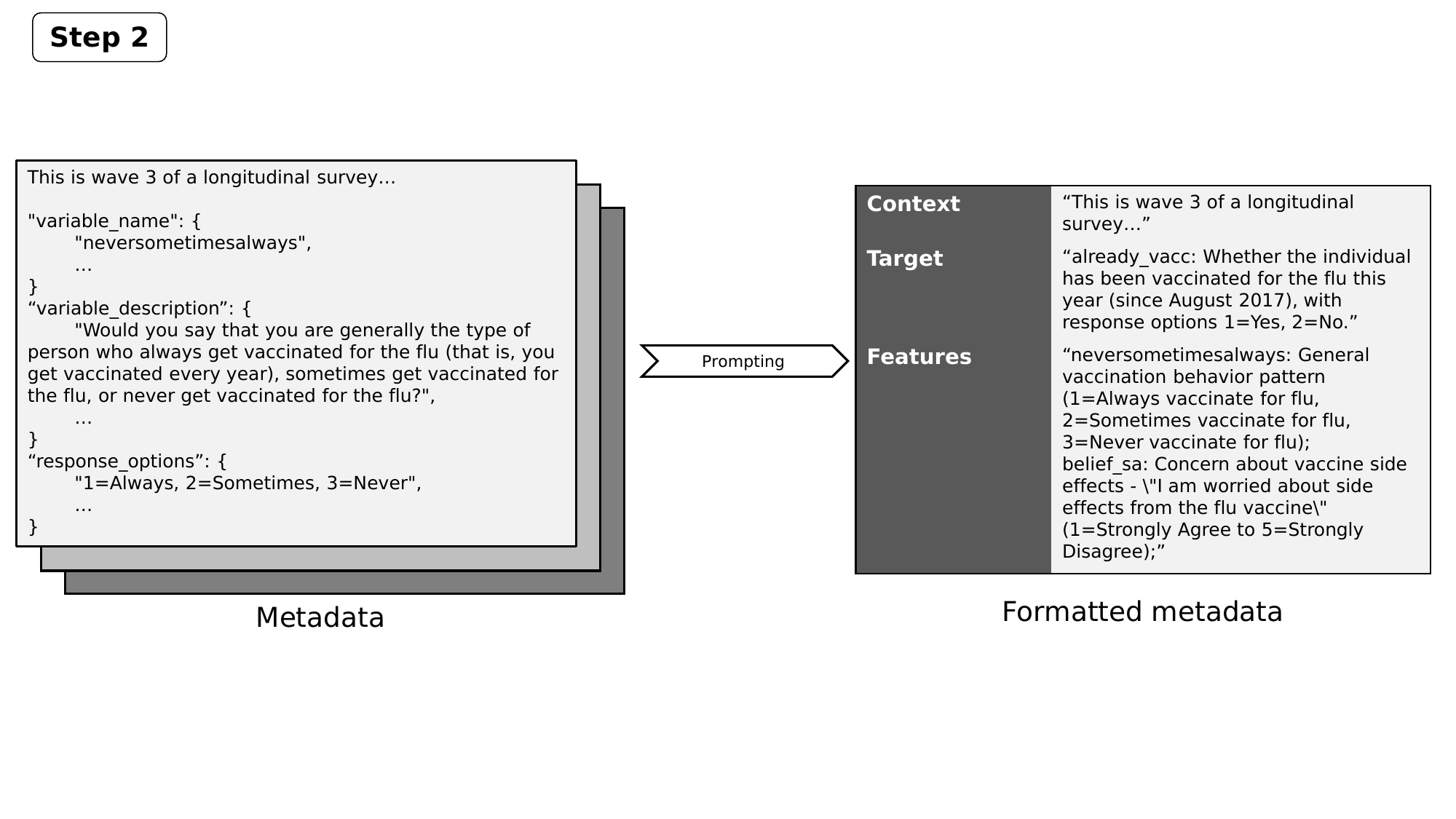}
\caption{Step 2: Formatted metadata generation. An example of metadata and formatted metadata.}
\label{fig:step2}
\end{figure}

We prompt the LLM using a cluster-informed approach similar to \cite{unipredict2024}, leveraging raw dataset metadata that includes the original survey questions, variable description, and the range of possible responses for each questions.
The objective of this step is to transform original metadata (inconsistently formatted) into standardized natural-language summarizes.
Specifically, we standardize dataset information into a structured format consisting of the dataset’s context, target variable, and feature meanings. 
This standardized representation improves the interpretability and consistency of the dataset descriptions across different waves and facilitates subsequent reasoning and analysis by the LLM.
Figure \ref{fig:step2} presents an example of the original metadata and its corresponding formatted version.
We refer readers to Appendix \ref{Sec:Appendix_Cluster_Prompt} for a more detailed version of metadata, survey questions and possible responses, the prompt text, and a sample output.

\subsection*{Step 3: Synthetic data prompting}
\begin{figure}[!ht]
\centering
\includegraphics[width=\linewidth]{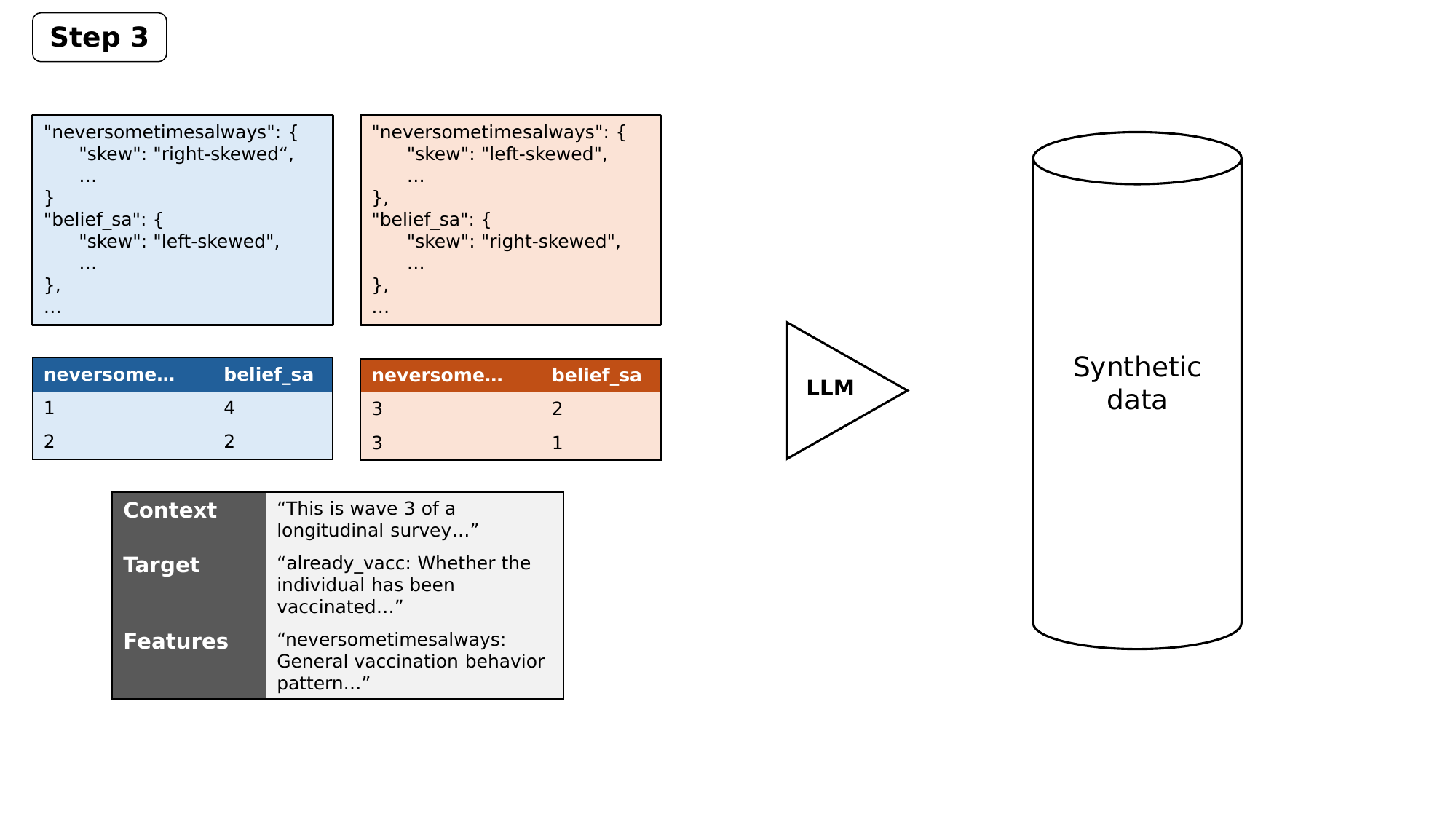}
\caption{Step 3: Synthetic data prompting. We uses the outputs produced in the previous stages of the workflow as inputs to prompt the synthetic data generation.}
\label{fig:step3}
\end{figure}

We prompt the LLMs for synthetic data generation using the outputs produced in the previous stages of the workflow. 
Specifically, we use the natural-language cluster descriptions and cluster samples in Step 1, together with the formatted metadata produced in Step 2, as inputs to the generation process, illustrated in Figure \ref{fig:step3}.
To examine the performance of the proposed workflow across different model families, we experiment with four LLMs: Claude Sonnet 4, Gemini 1.5 Pro, GPT 3.5 Turbo, and GPT 4o Mini.
We generate 100 outputs in each experimental run. 
Precisely, we prompt the same model 10 times, with the exact same input data, asking it to generate 10 synthetic data points.
We adapt this prompting method because directly prompting for 100 data points causes the model to output an inconsistent number of synthetic data points.
We generate 700 synthetic data points from each listed LLM.
For consistency, each execution of the workflow uses the same LLM throughout all stages of the pipeline.

\subsection*{Evaluation}
We evaluate the quality of the generated data by measuring its similarity in data spread and distribution shape compared to the original data. 
We adopt four metrics to evaluate data similarity, which are the \textit{shape}, the \textit{trend}, the \textit{Classifier Two-Sample Test (CS2T)}, and the \textit{overall} score. 
Those four metrics are widely used in the data generation community, see \cite{shi2025}. 
The \textit{shape} similarity score describes how similar individual variable distributions are in the synthetic data as compared to the real data. The higher the score, the more similar their distributions. A score is obtained for each feature, and the final \textit{shape} similarity score is the average cross all features.
The \textit{trend} similarity score describes how similar pairwise variable correlations are in the synthetic data as compared to the real data. The higher the score, the more similar the correlations. A \textit{trend} score is obtained for each feature, and the final \textit{trend} similarity score is the average of all features.
The Classifier Two-Sample Test \textit{(CS2T)} score indicates how difficult it is to tell apart the synthetic data from the real data. A score of 1 indicates that the model does not distinguish at all between the real and synthetic rows (i.e., does not do better than chance), while a score of 0 indicates that it perfectly distinguishes the real from the synthetic data.
The \textit{overall} score is computed by taking the average of the shape score and the trend score, which is a concise indicator of how well the synthetic data mimics the statistical characteristics of the real data.
All four scores lie between 0 and 1.

Additionally, we compare the vaccination rates estimated from the synthetic data with the true vaccination rates estimated by the real data in the FluPaths dataset. To estimate the true vaccination rate, we use the variables \texttt{already\_vaccinated\_flu}, which measures whether the respondent received the flu vaccine in the year that the survey was applied, and \texttt{already\_vaccinated\_flu\_rec}, which measures whether the respondent received the flu vaccination in response to a visit to a healthcare provider in the year that the survey was applied. For each respondent in each wave, we build another variable, \texttt{already\_vacc}, which pools the information from \texttt{already\_vaccinated\_flu} and \texttt{already\_vaccinated\_flu\_rec}. Namely, it will equal True if at least one of from \texttt{already\_vaccinated\_flu} and \texttt{already\_vaccinated\_flu\_rec} evaluates to True. If none of them evaluate to True, and at least one of them evaluates to False, \texttt{already\_vacc}will evaluate to False. Otherwise, both \texttt{already\_vaccinated\_flu} and \texttt{already\_vaccinated\_flu\_rec} are missing for the given respondent, and \texttt{already\_vacc} will be marked as missing as well. Then, for each wave, we estimate the vaccination rate in both the real and synthetic data as the number of participants for which \texttt{already\_vacc} equals True, divided by the total number of participants on each wave. 


\subsection*{Zero-shot and Few-shot Prompting}
We also compare our proposed cluster-informed prompting approach with zero-shot and few-shot baselines. 
Zero-shot prompting is a prompting technique in which a LLM is asked to perform a task without any task-specific examples in the prompt. The model relies entirely on its pretrained knowledge and the natural language instruction supplied by the user.
In contrast, few-shot prompting involves providing a small number of examples within the prompt before presenting the target task. In our experiments, we provide questions and responses from human participants in the FluPaths dataset. 
We include examples of zero-shot and few-shot prompts in Appendix \ref{Sec:zero_few}. 

\section{Results}

We first present metrics of data similarity of different LLMs across all all waves in Table~\ref{tab:results}.
We observe that the shape similarity scores are relatively stable, with all models scoring consistently around 0.8 in all waves. The exception of value drops to 0.68 for GPT 3.5 Turbo and GPT 4o Mini for Wave 1. 
The stable shape scores suggest that individual variable distributions in the real dataset are roughly preserved in the synthetic data.
The trend scores, on the other hand, tend to be lower, generally lying between 0.5 and 0.7. 
Hence, it is not clear whether correlations in the real dataset are preserved in the synthetic data.
The C2ST scores are generally below 0.5, and in fact most models yield scores around 0.35 or lower. 
This phenomenon indicates that synthetic data samples generated by LLMs are generally distinguishable from real data samples.
The exception is Gemini 1.5 Pro whose C2ST scores are consistently around or above 0.5.
Therefore, it suggests that Gemini 1.5 Pro produces synthetic data that is difficult to  distinguish from real data. 
The overall score is the average of the shape score and the trend score, which indicates how well the synthetic data mimics the statistical characteristics of the real data. 
We can conclude that Claude Sonnet 4 generated the highest quality synthetic data, in the sense that it exhibits the closest statistical characteristics to the real data.
In general, LLMs are able to capture the distributions of individual variable in the real data.
However, it remains unclear whether they can effectively preserve pairwise correlations.
Moreover, individual synthetic responses are generally distinguished from real responses.

\begin{table}[!htbp]
\centering
\small
\setlength{\tabcolsep}{5pt}
\renewcommand{\arraystretch}{0.95}

\begin{tabular}{llcccc}
\toprule
Wave & Model & Shape & Trend & C2ST & Overall \\
\midrule

\multirow{4}{*}{1}
& Claude Sonnet 4 & 0.780 & \textbf{0.623} & 0.360 & 0.701 \\
& Gemini 1.5 Pro & \textbf{0.807} & 0.620 & \textbf{0.625} & \textbf{0.714} \\
& GPT-3.5 Turbo & 0.682 & 0.513 & 0.361 & 0.597 \\
& GPT-4o Mini & 0.687 & 0.532 & 0.289 & 0.610 \\

\midrule

\multirow{4}{*}{3}
& Claude Sonnet 4 & \textbf{0.835} & 0.704 & 0.377 & \textbf{0.769} \\
& Gemini 1.5 Pro & 0.830 & 0.679 & \textbf{0.571} & 0.755 \\
& GPT-3.5 Turbo & 0.804 & \textbf{0.725} & 0.361 & 0.764 \\
& GPT-4o Mini & 0.802 & 0.686 & 0.159 & 0.744 \\

\midrule

\multirow{4}{*}{5}
& Claude Sonnet 4 & 0.848 & \textbf{0.717} & 0.315 & \textbf{0.782} \\
& Gemini 1.5 Pro & \textbf{0.870} & 0.688 & \textbf{0.461} & 0.779 \\
& GPT-3.5 Turbo & 0.788 & 0.672 & 0.181 & 0.730 \\
& GPT-4o Mini & 0.798 & 0.685 & 0.193 & 0.741 \\

\midrule

\multirow{4}{*}{7}
& Claude Sonnet 4 & \textbf{0.814} & \textbf{0.689} & 0.357 & \textbf{0.751} \\
& Gemini 1.5 Pro & 0.810 & 0.627 & 0.494 & 0.718 \\
& GPT-3.5 Turbo & 0.768 & 0.580 & 0.229 & 0.674 \\
& GPT-4o Mini & 0.793 & 0.672 & \textbf{0.518} & 0.732 \\

\bottomrule
\end{tabular}
\caption{Similarity scores for different LLMs across Waves 1, 3, 5, and 7.}
\label{tab:results}
\end{table}

\begin{figure}[!ht]
\centering
\includegraphics[width=0.7\linewidth]{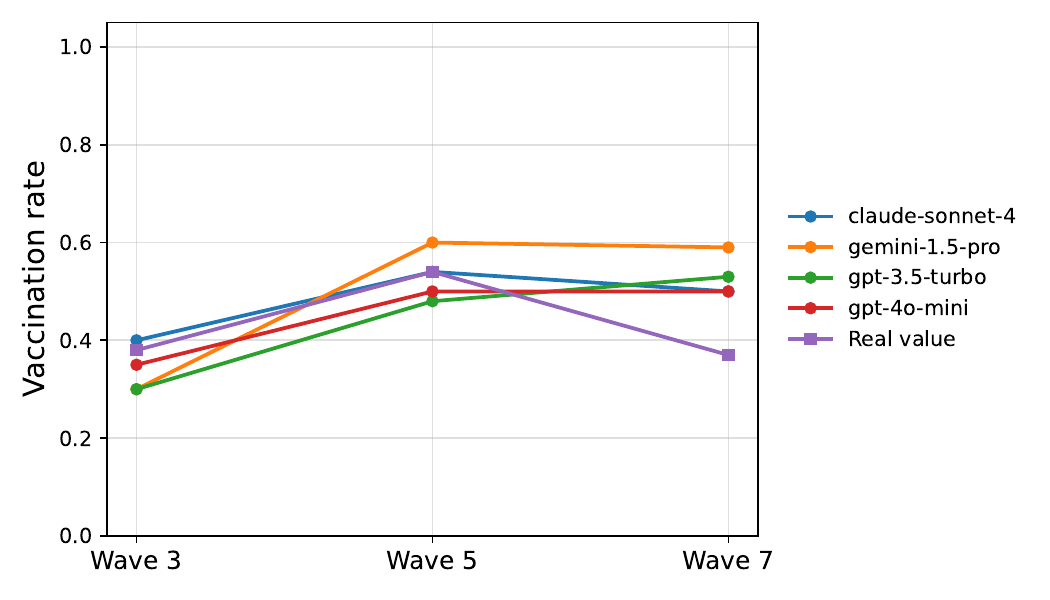}
\caption{Real and synthetic vaccination rates across waves.}
\label{fig:synthetic vaccination rates}
\end{figure}

We now compare the real and synthetic vaccination rates. We only consider Waves 3,5, and 7 here because the vaccination outcome variable is not included in Wave 1.
In Figure~\ref{fig:synthetic vaccination rates}, we depict the real vaccination rates and the synthetic vaccinate rates produced by four different LLMs.
We observe that all synthetic vaccination rates produced by LLMs generally follow the trend of the real vaccination rates across three waves.
However, all LLMs overestimate the vaccination rate in Wave 7.
Claude Sonnet 4 consistently produces estimates that are the closet to the real vaccination rates, whereas Gemini 1.5 Pro exhibits the least consistent trend relative to the real vaccination rates.
Additionally, the Appendix \ref{Sec:zero_few} shows additional results for the zero-shot and few-shot prompting baselines. 
Figures \ref{fig: distribtion_avarge_zero_shot} and \ref{fig: distribtion_avarge_few_shot} present the vaccination rates produced by various LLMs using the zeros-shot prompting approach, and few-shot prompting approach, respectively.
Both zero-shot and few-shot prompting show systematic overestimation of vaccination rates, suggesting that cluster-informed prompting provides meaningful gains over simpler approaches.

We then separately compare the real and synthetic vaccination rates within each cluster. 
For Waves 3 and 5, Claude Sonnet 4 overestimates Cluster 1 and underestimates Cluster 2, while GPT 3.5 Turbo does the opposite. 
GPT 4o Mini and Gemini 1.5 Pro follow a similar trend. In Wave 3, they underestimate both clusters, while they underestimate Cluster 1 but overestimate Cluster 2 in Wave 5. 
For Wave 7, all models overestimate the vaccination rates in both clusters.
GPT 4o Mini produces the estimate that most closely matches the real rates for Cluster 1, whereas Claude Sonnet 4 provides the closet estimate for Cluster 2.
Overall, it is unclear whether any model can consistently capture the cluster-specific vaccination rates, and we leave this as an important direction for future analysis.


\begin{table}[!b]
\centering
\small
\begin{tabular}{llccc}
\toprule
Cluster & Model & Wave 3 & Wave 5 & Wave 7 \\
\midrule

\multirow{5}{*}{Cluster 1}
& Real rate       & \textbf{0.469} & \textbf{0.651} & \textbf{0.494} \\
& Claude Sonnet 4   & 0.600 & 0.700 & 0.620 \\
& Gemini 1.5 Pro    & 0.392 & 0.640 & 0.600 \\
& GPT 3.5 Turbo     & 0.250 & 0.500 & 0.547 \\
& GPT 4o Mini       & 0.415 & 0.528 & 0.528 \\

\midrule

\multirow{5}{*}{Cluster 2}
& Real rate        & \textbf{0.299} & \textbf{0.418} & \textbf{0.273} \\
& Claude Sonnet 4   & 0.200 & 0.380 & 0.380 \\
& Gemini 1.5 Pro    & 0.204 & 0.560 & 0.580 \\
& GPT 3.5 Turbo     & 0.333 & 0.458 & 0.512 \\
& GPT 4o Mini       & 0.277 & 0.469 & 0.531 \\

\bottomrule
\end{tabular}

\caption{Synthetic vaccination rates by cluster across survey waves. Cluster 1 is characterized by positive attitudes toward vaccination, whereas Cluster 2 exhibits negative attitudes toward vaccination.}
\label{tab:syn_vacc_rates_per_cluster}
\end{table}

\section{Discussion}

In this work, we addressed the problem of augmenting survey data by leveraging large language models (LLMs) and the FluPaths and COVIDPaths datasets collected by the RAND Corporation. We performed dimension reduction and cluster analysis to help understand the pattern in our datasets and help inform our methodologies. We obtained groups aligning with generally positive and generally negative attitudes towards vaccination in Waves 1, 3, 5 and 7. We also performed a cross-wave analysis of these clustering results, and identified a group with generally stable positive attitudes towards vaccination, group with generally stable negative attitudes towards vaccination, and a "drifter" group, consisting of people that either changed their overall attitudes towards vaccination, or maintained generally neutral attitudes towards vaccination.

We explored a cluster-informed prompting methodology, in which we generated synthetic responses to a subset of questions on Waves 1, 3, 5 and 7, after incorporating results from our cluster analysis. We found that the resulting synthetic datasets captured individual variable distributions in the real data, but it is unclear whether pairwise correlations are captured. Additionally, individual synthetic responses are generally distinguishable from real responses. These remain as points to be addressed by future work. 


Several recent approaches illustrate directions for improving synthetic survey and behavioral data generation with LLMs. The HARMONIC method \citep{wang2024harmonic} generates synthetic tabular data by instruction fine-tuning LLMs on k-nearest-neighbor–constructed datasets that preserve inter-row relationships while permuting feature columns to prevent memorization. \cite{suh2025language} instead apply LoRA parameter-efficient tuning with a forward KL-divergence loss, penalizing underestimation of high-probability human responses and showing robustness across survey domains and demographic groups. Other work emphasizes inference-time strategies: \cite{huang2024fewer} propose diversifying prompts through task-agnostic and task-specific perturbations, reducing error rates and improving convergence in Best-of-N sampling, while \cite{li2025llms} introduce Generative Self-Aggregation, a two-stage prompting framework that synthesizes multiple candidate responses into a higher-quality aggregated solution. Together, these methods highlight complementary directions in fine-tuning, loss design, and prompting that can inform future efforts to align LLMs with human behavioral data.


Recent commercial efforts, such as Expected Parrot \citep{Horton2024EDSL}, have begun offering pipelines that create LLM-based survey respondents, highlighting the growing practical interest in human behavior simulation. Their approach primarily relies on prompt-based profile template generation, where demographic and persona descriptions are used to condition pretrained language models during survey response generation.

Future work could explore counterfactual scenario generation, in which agent-based model states at specific time steps are fed to fine-tuned LLMs to test whether the model can replicate contemporaneous survey responses, closing the loop between simulation and synthetic behavioral data. 



\section{Acknowledgments}
This work was conducted as part of the Research in Industrial Projects for Students (RIPS) program, hosted by the Institute for Pure and Applied Mathematics (IPAM) at the University of California, Los Angeles. We sincerely thank Dr.~Christian Ratsch, Dr.~Dimitri Shlyakhtenko, and Dr.~Susana Serna for organizing and supporting the RIPS program, through which this collaboration was initiated and developed.

We also gratefully acknowledge the support of IPAM and the National Science Foundation (NSF). IPAM, including the RIPS program, is supported through NSF funding.

The FluPaths and COVIDPaths longitudinal surveys that form the empirical basis of this work were developed and conducted with support from the National Institute of Allergy and Infectious Diseases (NIAID) of the National Institutes of Health under Award Number R01AI160240. The content is solely the responsibility of the authors and does not necessarily represent the official views of the National Institutes of Health.

The authors additionally thank the participants of the American Life Panel, provided by the RAND Corporation, for their continued engagement across survey waves. We also acknowledge Dr.~Andrew M. Parker, Dr.~Ashley Gromis, Mr.~Dulani Woods, and Ms.~Katrina Rivera for their contributions to the underlying data pipeline and survey infrastructure.

\bibliographystyle{abbrvnat}
\bibliography{references}

\clearpage
\appendix
\section{Supplemental Material}
In this appendix, we provide supplemental templates for cluster-based prompting techniques described in Section \ref{Sec: method}, and for zero-shot and few-shot prompting baselines.

\subsection{Cluster-based Prompting}
\label{Sec:Appendix_Cluster_Prompt}

The workflow of cluster-based prompting approach described in Section~\ref{Sec: method} involves three steps. Here we include the prompts we use.

In \textbf{Step 1}, for each feature $X$ and a cluster $C$, we prompt the LLM with summary statistic of $X$ conditioned to $C$. In the following prompt, \textit{var\_type} is either ``ordinal'' or ``binary'' depending on the responses to the corresponding question. We use skew, spread and shape given that these are qualitative ways to describe a distribution while helping prevent the LLM in Step 3 from merely memorizing summary statistics. Therefore, the description of each cluster consists of a qualitative description of each selected variable in the original wave dataset.
We repeat the following prompts for each selected variable to obtain each cluster's description.

\begin{small}
\begin{verbatim}
You are an expert at giving short qualitative descriptions of variable
distributions. Please describe the distribution of {var_type} variable that
takes values from {min} to {max} and has the following summary statistics: 
Min: {min}
First quartile: {1q}
Median: {median}
Mean: {mean}
Third quartile: {3q}
Max: {max}
Variance: {variance}
Do not use any numbers in your answer. Respond, in at most one sentence, 
to the questions below, and only those questions. Be succinct in your answers.
Skew:
Spread:
Shape:
\end{verbatim}
\end{small}

Below is an example of sample output produced by a large language model.
\begin{small}
\begin{verbatim}
"howlongagovaccine": {
    "skew": "highly right-skewed",
    "spread": "low variability",
    "shape": "heavily concentrated at the minimum value"
}
\end{verbatim}
\end{small}

In \textbf{Step 2}, we prompt the LLM with the metadata to generate a natural language description for this metadata in a standard format. The prompt has two inputs:
\begin{itemize}
\item A short description of the dataset and its purpose. Example:
\begin{small}
\begin{verbatim}
This is wave 3 of a longitudinal survey studying behavioral and beliefs
aspects related to influenza, including, but not limited to, healthcare
behaviors and risk perceptions. The target variable is already_vacc
(whether the individual already vaccinated). This survey was conducted in
Fall 2017. The previous wave was conducted in Fall 2016.
\end{verbatim}
\end{small}
\item Survey questions and possible responses. Example:
\begin{small}
\begin{verbatim}
    {
    "variable_name": {
        "0": "calcage",
        "1": "gender",
        "2": "neversometimesalways",
        "3": "belief_xn",
        ...
    },
    "variable_description": {
        "0": "What is your age?",
        "1": "What is your gender?",
        "2": "Would you say that you are generally the type of person who
        always get vaccinated for the flu (that is, you get vaccinated every
        year), sometimes get vaccinated for the flu, or never get vaccinated 
        for the flu?",
        "3": "I am afraid of the needles used for vaccination.",
        ...
    },
    "response_options": {
        "0": "Integer",
        "1": "1=1 Male, 2=2 Female",
        "2": "1=1 Always vaccinate for flu, 2=2 Sometimes vaccinate for flu,
        3=3 Never vaccinate for flu",
        "3": "1=1 Strongly Agree, 2=2 Agree, 3=3 Neither Agree nor Disagree,
        4=4 Disagree, 5=5 Strongly Disagree".
        ...
    }
}
\end{verbatim}
\end{small}
\end{itemize}

The following is the prompt text:

\begin{small}
\begin{verbatim}
You are an expert at describing datasets succinctly.The following is the
metadata of a tabular dataset. {wave_context}
Column meanings are as follows: {wave_dict}

Instructions:

Generate a description of the given dataset containing the following:
1. The dataset description, in one sentence.
2. The target of the dataset. If no target exists, choose one from the column
as target for the dataset to classify.
3. The features and their explanations.
\end{verbatim}
\end{small}

Below is a sample output produced by a large language model.
\begin{small}
\begin{verbatim}
{
  "context": "This is wave 3 of a longitudinal survey studying behavioral and
  beliefs aspects related to influenza vaccination, conducted in Fall 2017
  following a previous wave in Fall 2016.",
  "target": "already_vacc: Whether the individual has been vaccinated for the
  flu this year (since August 2017), with response options 1=Yes, 2=No.",
  "feature_meanings": "calcage: The individual's age (integer value); gender:
  The individual's gender (1=Male, 2=Female); neversometimesalways: General
  vaccination behavior pattern (1=Always vaccinate for flu, 2=Sometimes 
  vaccinate for flu, 3=Never vaccinate for flu); belief_xn: Belief about 
  vaccination needles - \"I am afraid of the needles used for vaccination\" 
  (1=Strongly Agree to 5=Strongly Disagree); belief_sa: Concern about vaccine 
  side effects - \"I am worried about side effects from the flu vaccine\" 
  (1=Strongly Agree to 5=Strongly Disagree); cluster: General trend of the 
  individual's current opinion about vaccination (1=Roughly positive, 2=Roughly 
  negative); prev_cluster: General trend of the individual's opinion about 
  vaccination in the previous flu season (1=Roughly positive, 2=Roughly 
  negative)."
}
\end{verbatim}
\end{small}

In \textbf{Step 3}, we prompt the LLMs using cluster examples (here labeled \textit{examples\_cluster1} and \textit{examples\_cluster2}) and the cluster descriptions from Step 1, together with the formatted metadata from Step 2 (here labeled \textit{wave\_metadata}). We also specify the desired number of output synthetic data points (labeled \textit{n\_outputs}). 

\begin{small}
\begin{verbatim}
You are an epidemiological data simulation expert that generates realistic data
in JSON format. Below you can find the metadata of a dataset, which contains 
the context, target and feature meanings. {wave_metadata}
Clusters were found in the dataset. The distribution of the variables in each 
cluster can be described as follows: {cluster_descriptions}

Below are two examples from the first cluster: {examples_cluster1}
Below are two examples from the second cluster: {examples_cluster2}

INSTRUCTIONS:
1. Generate data for exactly {n_outputs} individuals conforming to the given
metadata and examples.
2. Use the provided examples and cluster descriptions to recreate those 
patterns to the best of your ability.
3. Consider real-world factors, including, but not limited to: vaccine 
hesitancy, changing attitudes, seasonal patterns, demographics, and health 
status.
\end{verbatim}
\end{small}


Below we include the first five rows of output of Step 3 for Wave 3.

\begin{table}[!htbp]
\centering
\resizebox{\textwidth}{!}{%
\begin{tabular}{@{}llllllll@{}}
\toprule
\texttt{calcage} & \texttt{gender} & \texttt{neversometimesalways} & \texttt{belief\_xn} & \texttt{belief\_sa} & \texttt{already\_vacc} & \texttt{cluster} & \texttt{prev\_cluster} \\ \midrule
45      & 2      & 1                    & 5         & 4         & 1            & 1       & 1            \\
67      & 1      & 1                    & 4         & 5         & 1            & 1       & 1            \\
29      & 2      & 2                    & 3         & 4         & 2            & 1       & 1            \\
52      & 1      & 1                    & 5         & 4         & 2            & 1       & 2            \\
38      & 2      & 1                    & 4         & 3         & 1            & 1       & 1            \\ \bottomrule
\end{tabular}%
}
\end{table}

\subsection{Zero-shot and Few-shot prompting}
\label{Sec:zero_few}
In this section, we include the prompts for zero-shot and few-shot prompting, along with the corresponding figures showing the estimated vaccination rates produced by those two approaches. 
We compare them with the estimated vaccination rates from the cluster-based prompting in Section \ref{Sec: method}.
\subsubsection{Zero-shot Prompting}

We present an example of prompts for zero-shot prompting technique. 

\begin{small}
\begin{verbatim}
You are tasked with generating realistic synthetic data for flu vaccination 
patterns. Your goal is to create data that closely resembles real-world 
epidemiological patterns and human decision-making behaviors regarding flu 
vaccination.

TASK: Generate vaccination decision data for 10 different people across 8 flu 
seasons (2016-2017 through 2023-2024).

CONTEXT: Consider these real-world factors that influence flu vaccination 
decisions:
- Age demographics (older adults more likely to vaccinate)
- Health conditions (chronic conditions increase vaccination likelihood)
- Previous vaccination history (people tend to be consistent in their behavior)
- Seasonal variations (some years have higher vaccination rates due to severe 
  flu outbreaks)
- COVID-19 pandemic impact (2020-2021 and 2021-2022 seasons saw changes in 
  healthcare-seeking behavior)
- Access to healthcare and socioeconomic factors
- Personal beliefs and risk perception

SEASONS TO INCLUDE:
2016-2017, 2017-2018, 2018-2019, 2019-2020, 2020-2021, 2021-2022, 2022-2023, 
2023-2024

OUTPUT FORMAT REQUIREMENTS:
For each of the 10 people, provide:
1. A brief demographic/health profile explaining their vaccination tendencies
2. Vaccination decision for each season (1 = vaccinated, 0 = not vaccinated)

Structure your response as follows:

Person 1:
Profile: [Brief explanation of demographics, health status, and general 
         vaccination tendency - 1-2 sentences]
Decisions: 2016-2017: [0 or 1], 2017-2018: [0 or 1], 2018-2019: [0 or 1], 
          2019-2020: [0 or 1], 2020-2021: [0 or 1], 2021-2022: [0 or 1], 
          2022-2023: [0 or 1], 2023-2024: [0 or 1]

Person 2:
[Continue for all 10 people...]

IMPORTANT: Make the data realistic by:
- Varying vaccination rates across different demographic profiles
- Showing some consistency in individual behavior over time
- Reflecting known epidemiological patterns (e.g., higher rates in elderly, 
  during severe flu years)
- Accounting for the COVID-19 pandemic's impact on healthcare behaviors in 
  2020-2022
- Including a mix of consistent vaccinators, non-vaccinators, and those who 
  vary year to year

Generate diverse, realistic profiles that represent different segments of the 
population.
\end{verbatim}
\end{small}

\begin{figure}[!htbp]
\centering
\includegraphics[width=0.9\linewidth]{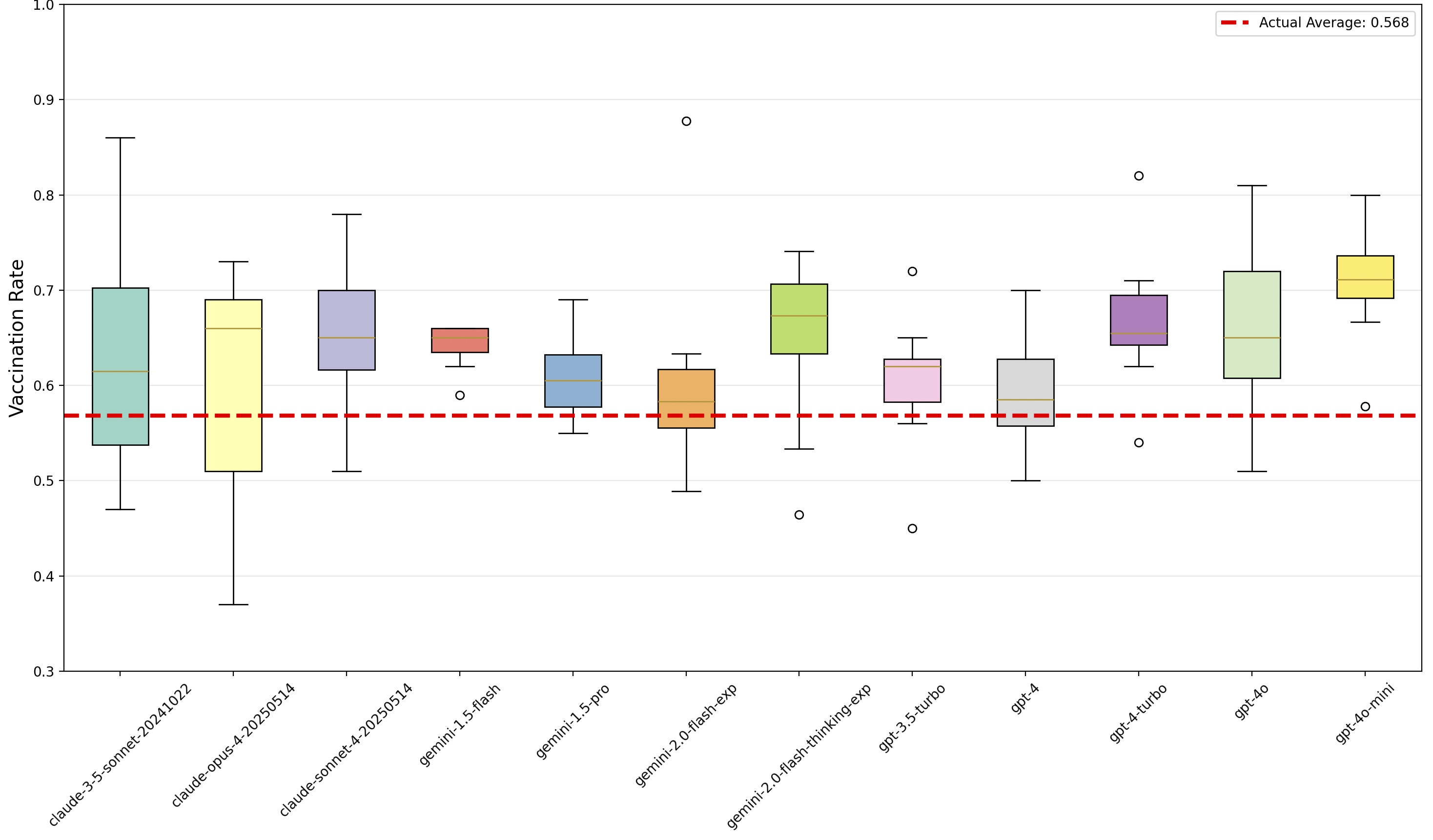}
\caption{Vaccination rates generated by LLMs using the zero-shot prompting approach.}
\label{fig: distribtion_avarge_zero_shot}
\end{figure}

In Figure \ref{fig: distribtion_avarge_zero_shot}, we show vaccination rates generated by various LLMs using the zero-shot prompting approach.
Each box represents the distribution of average vaccination rates across all iterations for a given model. The red dashed line indicates the actual average vaccination rate in the real dataset (0.568). We observe that all LLMs overestimate the vaccination rate. 

\subsubsection{Few-shot Prompting}

We present an example of prompts for few-shot prompting technique. 

\begin{small}
\begin{verbatim}
You are tasked with generating realistic synthetic data for flu vaccination 
patterns using few-shot learning. You have been provided with real examples 
from human participants across different flu seasons.

IMPORTANT: These examples are provided as LEARNING OPPORTUNITIES to understand 
patterns and relationships - DO NOT simply copy or replicate the exact patterns 
you see. Instead, use them to understand the underlying factors that influence 
vaccination decisions and create NEW, realistic synthetic data that reflects 
similar decision-making processes but with your own variation.

TASK: Generate vaccination decision data for 10 different people across 8 flu 
seasons (2016-2017 through 2023-2024), based on the patterns you observe in 
the real examples below.

REAL HUMAN EXAMPLES WITH PREDICTOR CORRELATIONS:

=== [SEASON NAME] ===

PREDICTOR VARIABLES AND THEIR CORRELATIONS WITH VACCINATION:
  • Variable Name: Question text
    Correlation with vaccination: [correlation] (Positive/Negative relationship)

REAL HUMAN EXAMPLES:

Example 1:
  Q: [Question text]
  A: [Answer] (Positive/Negative)
  VACCINATION DECISION: [Decision] (Vaccinated/Not Vaccinated)

CRITICAL INSTRUCTIONS:
DO NOT COPY THE PATTERNS: Use the examples above to UNDERSTAND 
    decision-making factors, but create your own realistic variations
LEARN FROM CORRELATIONS: Pay attention to how different predictor 
    variables correlate with vaccination decisions
CREATE DIVERSITY: Generate varied profiles that reflect different 
    demographic groups and decision-making styles
MAINTAIN REALISM: Ensure your synthetic data reflects plausible 
    human behavior patterns

ANALYSIS REQUIREMENTS:
Based on the real human examples and correlation patterns above, generate 
data for 10 NEW synthetic participants. For each participant, provide:

1. PARTICIPANT PROFILE: Brief demographic/health profile explaining their 
   vaccination tendencies (1-2 sentences)

2. VACCINATION DECISIONS: Decision for each season (1 = vaccinated, 
   0 = not vaccinated)
   Format: 2016-2017: [0 or 1], 2017-2018: [0 or 1], ..., 2023-2024: [0 or 1]

3. OVERALL REASONING: Your overall reasoning and justification for the 
   patterns you created

4. FEW-SHOT INSIGHTS: How the few-shot examples helped you understand 
   vaccination behavior patterns

5. DECISION INFLUENCE: How the real examples influenced your synthetic 
   data generation decisions

6. ADDITIONAL DATA NEEDS: What other data you would want in few-shot 
   examples to generate more accurate responses

7. CONFIDENCE SCORE: Your confidence in the accuracy of your response 
   (scale 1-10, where 10 = very confident)

OUTPUT FORMAT:
Person 1:
Profile: [Brief explanation]
Decisions: 2016-2017: [0 or 1], 2017-2018: [0 or 1], 2018-2019: [0 or 1], 
          2019-2020: [0 or 1], 2020-2021: [0 or 1], 2021-2022: [0 or 1], 
          2022-2023: [0 or 1], 2023-2024: [0 or 1]

[Continue for all 10 people...]

Overall Reasoning: [Your detailed reasoning]

Few-Shot Insights: [How examples helped understanding]

Decision Influence: [How examples influenced decisions]

Additional Data Needs: [What other data would be helpful]

Confidence Score: [1-10]

REMEMBER: The goal is to create realistic synthetic data that captures 
similar decision-making patterns to the real examples, but with your own 
creative and realistic variations. Do not simply replicate what you see.
\end{verbatim}
\end{small}

\begin{figure}[!ht]
\centering
\includegraphics[width=0.85\linewidth]{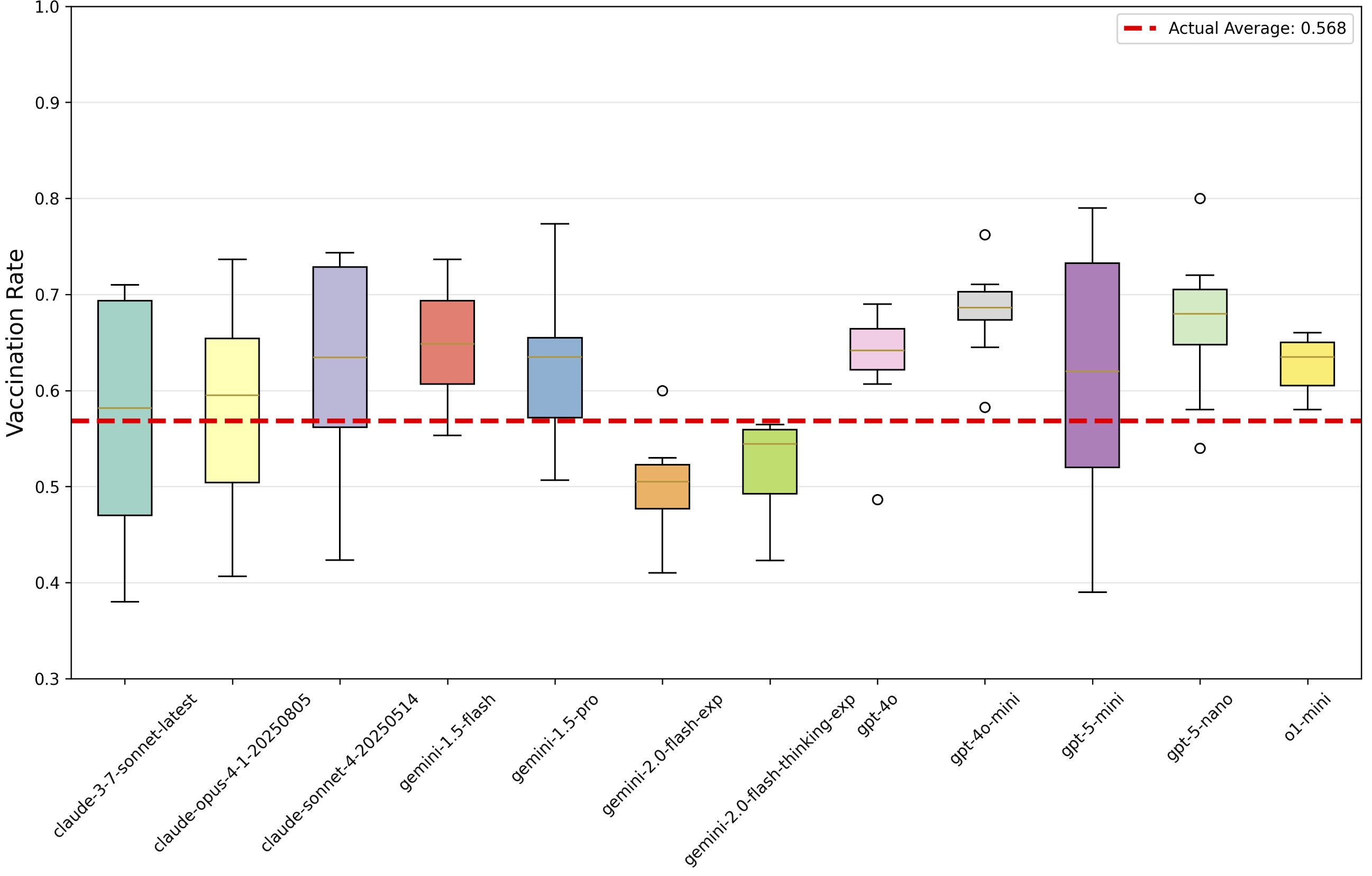}
\caption{Vaccination rates generated by LLMs using the few-shot prompting approach.}
\label{fig: distribtion_avarge_few_shot}
\end{figure}

In Figure \ref{fig: distribtion_avarge_few_shot}, we show vaccination rates generated by various LLMs using the few-shot prompting approach.
Each box represents the distribution of average vaccination rates across all iterations for a given model. The red dashed line indicates the actual average vaccination rate in the real dataset (0.568). We observe that most LLMs overestimate the vaccination rate. 

\end{document}